\documentclass[aps,prb,preprint]{revtex4}
\usepackage{graphicx}
\usepackage{amsmath}
\usepackage{amssymb}
\usepackage{ulem}

\begin{document}

\baselineskip=1.5cm

\title{Lattice-mismatch-induced granularity in CoPt-NbN and NbN-CoPt superconductor-ferromagnet heterostructures: Effect of strain.}

\author{R. K. Rakshit, S. K. Bose, R. Sharma, N. K. Pandey, and R. C. Budhani}

\email{rcb@iitk.ac.in}

\affiliation{Condensed Matter - Low Dimensional Systems
Laboratory, Department of Physics, Indian Institute of Technology
Kanpur, Kanpur - 208016, India}

\begin{abstract}

The effect of strain due to lattice mismatch and of ferromagnetic
(FM) exchange field on superconductivity (SC) in NbN-CoPt bilayers
is investigated. Two different bilayer systems with reversed
deposition sequence are grown on MgO (001) single crystals. While
robust superconductivity with high critical temperature (T$_c$
$\approx$ 15.3 K) and narrow transition width ($\Delta$T$_c$
$\approx$ 0.4 K) is seen in two types of CoPt-NbN/MgO
heterostructures where the magnetic anisotropy of CoPt is in-plane
in one case and out-of-plane in the other, the NbN-CoPt/MgO system
shows markedly suppressed SC response. The reduced SC order
parameter of this system, which manifests itself in T$_c$,
temperature dependence of critical current density J$_c$ (T), and
angular ($\phi$) variation of flux-flow resistivity $\rho$$_f$ is
shown to be a signature of the structure of NbN film and not a
result of the exchange field of CoPt. The $\rho$$_f$ (H,T,$\phi$)
data further suggest that the domain walls in the CoPt film are of
the N$\acute{e}$el type and hence do not cause any flux in the
superconducting layer. A small, but distinct increase in the
low-field critical current of the CoPt-NbN couple is seen when the
magnetic layer has perpendicular anisotropy.
\end{abstract}

\maketitle
\section{Introduction}
Hybrid superconductor-ferromagnet (SC-FM) structures have
generated a considerable amount of interest in recent years as
these provide model systems to understand the antagonism between
superconductivity and ferromagnetism
\cite{Ryazanov,Volkov,Buzdin,Allsworth,Rcb,Palau,Bulkaevskii,Cieplak,Gillijns,Jan,Kinsey,Lange,Lange1,Manios,Martin,Montero,Santiago,Gu,Yu}.
A rich variety of phenomena such as $\pi$-phase shift
\cite{Ryazanov,Buzdin}, triplet pairing \cite{Volkov,Buzdin},
field enhanced superconductivity \cite{Gu,Lange1}, domain wall
superconductivity \cite{Gillijns} and enhanced flux pinning
\cite{Montero,Cieplak,Lange,Santiago,Jan} etc. have been reported
in SC-FM structures. An important issue that needs to be
considered while addressing the physics of SC-FM-SC and FM-SC-FM
heterostructures is the difference in the crystallographic
structure and the degree of strain in the top and bottom SC or FM
layers, and interdiffusion at interfaces which could impart
different physical properties to the top and bottom layers. The
other issue related to the physics of SC-FM junctions is the
choice of materials. Thus far most of the studies in this area
have been carried out on heterostructures of elemental
superconductors such as Pb and Nb made in conjunction with 3d
transition metal ferromagnets
\cite{Lange1,Gillijns,Montero,Kinsey,Martin,Palau}. It is expected
that the competition between the SC and FM orders will have a
different flavor if hard superconductors characterized by a short
coherence length and large magnetic penetration depth are used.
The nature of magnetic layers is also not less important as field
enhanced superconductivity and vortex pinning depend greatly on
the domain structure and dynamics of domain wall motion in the
magnetic layer \cite{Lange}.

Here we report investigations of domain wall superconductivity,
flux pinning and granularity issues in a strongly type-II
superconductor placed in close proximity of a magnetic thin film.
The system investigated consists of a bilayer of A1 phase (fcc) or
L1$_0$ phase (fct) CoPt ferromagnet and NbN (rocksalt)
superconductor grown on MgO (001). NbN and CoPt were chosen as
constituents of the SC-FM bilayer since their respective
superconducting and magnetic properties can be modified in a
controlled way by altering the deposition conditions
\cite{Senapati,Budhani,Rakshit}. Two different bilayer systems
with reversed deposition sequence have been investigated to
address superconductivity and magnetism vis-a-vis crystallographic
structure and exchange field-induced proximity effects. We note
that while the lattice mismatch has a minor effect on
superconductivity of the bottom NbN layer in CoPt-NbN/MgO system,
in the NbN-CoPt/MgO bilayer it makes the NbN film to adopt a
strain-induced granular structure, which has interesting
repercussions on its superconducting response. We see a small but
distinct enhancement in the low-field critical current density
(J$_c$) of the CoPt-NbN/MgO bilayer when the CoPt has
perpendicular magnetic anisotropy.

\section{Experimental}

The CoPt-NbN/MgO and NbN-CoPt/MgO bilayers with 50 nm thick CoPt
and NbN layers in each case were deposited on (001) MgO in the
temperature range of 600 to 700 $^0$C by pulsed laser ablation of
Nb and CoPt targets in 99.9999 $\%$ pure nitrogen environment. A
distinctly different magnetic state of the CoPt is realized when
the deposition temperature is elevated from 600 to 700 $^0$C. At
the lower temperature, the CoPt is in the disordered fcc phase
(A1) with in-plane magnetization whereas, at 700 $^0$C it acquires
the ordered L1$_0$ structure with out-of-plane magnetic anisotropy
\cite{Barmak}. Further details of CoPt and NbN growth have been
reported earlier \cite{Budhani,Rakshit, Senapati}. The crystal
structure of bilayer films was investigated by X-ray diffraction
(XRD) measurements performed on a $\theta$ - $\omega$
diffractometer with CuK$_{\alpha1,\alpha2}$ radiation. The surface
topography of the bilayers was analyzed using high resolution
scanning electron microscopy (SEM), which showed a smooth
morphology for the CoPt-NbN/MgO system but a granular structure
with typical feature size of 75 to 100 nm in NbN-CoPt/MgO bilayer.
Measurements of AC susceptibility were performed with a
Hall-probe-based ac susceptometer, which is described in detail
elsewhere \cite{RSIkartik}. For measurements of magnetoresistance
and critical current density (J$_c$), samples were processed using
standard photolithography and Ar$^+$ ion milling to produce 500
$\mu$m long and 160 $\mu$m wide bridges. After removing the
photoresist, silver pads for electrical contact were deposited in
a four-probe configuration by thermal evaporation. The resistance
of the samples in the flux flow regime was also measured as a
function of the angle $\phi$ between the applied magnetic field
and film normal. While the angle in these measurements was changed
in a step of 2 degree, the field always remained orthogonal to the
current direction. A commercial magnetometer (Quantum Design
MPMS-XL5) was used for measurements of in-plane and out-of-plane
isothermal magnetization above and below the superconducting
transition temperature of the bilayers.

\section{Results and Discussion}
\subsection{Bilayers with in-plane magnetic anisotropy}
In Fig. 1(a) we have plotted the temperature dependence of
resistance of the two bilayers, deposited at 600 $^0$C along with
data for a pure NbN film for comparison. The latter shows a sharp
transition ($\Delta$T$_c$ $\approx$ 0.4 K) with the onset of
superconductivity at 15.5 K. The CoPt-NbN/MgO bilayer also
displays a sharp transition of width $\approx$ 0.4 K with T$_c$
onset at 15.3 K. The other bilayer (NbN-CoPt/MgO), however, has a
substantially reduced T$_c$ ($\approx$ 13.7 K) and a sizeable
broadening of the transition ($\approx$ 1.4 K).

Fig. 1(b) shows the real and imaginary components of the
fundamental susceptibility of the same samples. The frequency, and
amplitude of the ac field applied perpendicular to the plane of
the film, in these experiments are 121 Hz and 1 Oe, respectively.
The real part $\chi^{'}$ (T) of the complex susceptibility $\chi$
(T) = $\chi^{'}$ (T) + i$\chi^{''}$ (T) reflects the strength of
the induced shielding currents, while the imaginary part
$\chi^{''}$ (T) is connected to energy dissipation in the
material. The drop of $\chi^{'}$ (T) whose onset corresponds to
the temperature where the resistance (R) goes to zero is
reasonably sharp for both pure NbN film as well as for the
CoPt-NbN/MgO bilayer, indicating a homogeneous superconductor.
However, for the bilayer where the NbN is deposited on top of
CoPt, both $\chi^{'}$ and $\chi^{''}$ show significant broadening.
For inhomogeneous superconductor with distinctly different
intergrain and intragrain critical currents, the $\chi^{''}$ peak
generally splits into two components each corresponding to peak
dissipation in superconducting grains and in intergranular
material \cite{Classen}. The fact that we do not see two distinct
peaks in the $\chi^{''}$ data for NbN-CoPt/MgO system suggests
insignificant suppression of the order parameter in the intergrain
material of the NbN layer. However, the overall suppression of
T$_c$ seen here compared to the T$_c$ of the bilayer where NbN is
deposited first, seems to indicate a perturbation of the
electronic structure by stress or chemical doping which lowers the
critical temperature. In order to address these issues, we have
undertaken X-ray diffraction studies of the bilayer films.

Fig. 2 (Panel `a' and `b' respectively) shows the diffraction
profiles of NbN-CoPt/MgO and CoPt-NbN/MgO along with the profiles
of single layer CoPt (panel `c'), NbN (panel `d') and bare MgO
substrate (panel `e'). The single layer CoPt film shows only one
peak at 2$\theta$ = 48.13$^0$ which corresponds to the (200)
reflection of the disordered A1 (fcc) phase. Single layer NbN also
shows the characteristic (200) and (400) reflections of the fcc
phase indicating a highly textured growth along [001] direction of
MgO. While all these characteristic reflections of the [001]
growth are seen in CoPt-NbN/MgO bilayer (panel `b') as well, the
NbN peaks are not discernible in the reverse bilayer geometry
(panel `a'), where NbN was grown on CoPt, suggesting a granular
structure of the nitride. These results reveal that film growth
dynamics is greatly controlled by the amount of interfacial strain
induced by the cubic (100) MgO whose lattice parameter `a' is 4.21
{\AA}. NbN is also cubic with a = 4.39 {\AA}. The magnetic alloy
CoPt, when grown on (100) MgO at $\leq$ 600 $^0$C has a disordered
fcc structure with a lattice parameter of 3.772 {\AA}
\cite{Pennison}. The lattice mismatch for epitaxial growth of `A'
on `B' can be characterized in terms of the strain parameter
$\epsilon$ defined as $ \epsilon {\rm{ }} = {\rm{ }}{{d_{A}  -
d_{B} } \over {d_{A} }}{\rm{ }} \times {\rm{ }}100 $ where, d$_A$
and d$_B$ are the lattice parameters of `A' and `B' respectively.
This gives $\epsilon$ of $\approx$ 4.1 $\%$, $\approx$ - 11.6
$\%$, $\approx$ - 16.4 $\%$ and $\approx$ 14.1 $\%$ for NbN-MgO,
CoPt-MgO, CoPt-NbN and NbN-CoPt systems respectively. In Fig. 3 we
sketch the stacking of the unstrained unit cells in NbN, CoPt and
MgO starting with MgO at the bottom. Dotted lines in the figure
represents ideal coherent epitaxy in conformity with the
substrate. The origin of in-plane tensile strain on CoPt and
compressive strain on NbN under coherent epitaxy can be visualized
from this figure.

It is interesting to note that in spite of a large strain
parameter ($\epsilon$ $\approx$ - 16.4 $\%$), CoPt prefers to grow
epitaxially with a slight change ($\approx$ 0.2 $\%$) in lattice
parameter from bulk when deposited on NbN [Fig. 2(b)]. The peak
shift [$\Delta$$\theta$ $\approx$ 0.1$^0$] of the (200) reflection
of CoPt grown on NbN/MgO can be attributed to an improper growth
due to mismatch. On the other hand, the large compressive strain
($\sim$ 14.1 $\%$) experienced by NbN, when deposited on CoPt as
compared to only $\approx$ 4.1 $\%$ when grown directly on MgO,
forces a highly disordered growth of NbN. We believe that the
rough surface texture of CoPt films on MgO as seen in our scanning
electron microscopy studies and originating presumably from strain
($\epsilon$ $\approx$ - 11.6 $\%$) makes the subsequent growth of
NbN disordered.

In Fig. 4 (a $\&$ b) we have shown isothermal magnetization at 20
K as a function of in-plane magnetic field for the bilayers, along
with the data for resistance. We note that at 20 K the coercive
field (H$_c$) of the CoPt-NbN/MgO system is $\approx$ 250 Oe,
which is lower than the H$_c$ of the NbN-CoPt/MgO bilayer
($\approx$ 500 Oe) as seen in the M-H loop of Fig. 4(b). This can
be understood in the following way. During the growth of the 50 nm
thick NbN using the deposition conditions mentioned before, the
bottom CoPt layer gets annealed for roughly about 7 - 8 minutes,
which leads to a better ordering of the structure as evidenced by
the enhanced intensity of (200) reflection of the A1 phase (see
Fig.2 patterns `a' and `b'). This observation is consistent with
previous studies on CoPt system which reveal that the 600 $^0$C
deposited films consist of only the fcc phase and their coercive
field increases with the duration of postdeposition annealing
\cite{Budhani}.

The isothermal resistance of the samples in the normal state (20
K) when a dc field aligned parallel to the plane of the film and
directed perpendicular to current was scanned between + 3.5 kOe
and - 3.5 kOe is also shown in Fig. 4. Arrows in the figure mark
the increasing and decreasing branches of the field. For the
CoPt-NbN/MgO bilayer (Fig. 4(a)) upon reducing the field from 3.5
kOe, where the magnetization reaches saturation and the sample is
presumably in a single domain state, the resistance decreases
smoothly with field, i.e., dR/dH is positive. This behavior gives
way to a sudden jump in resistance when the sample reaches a
truely demagnetized state at H = H$_c$. We attribute this effect
to enhanced domain wall scattering of charge carriers which
presumably also leads to the observed superlinear dependence of R
on the increasing $\left|{H}\right|$ branch till the saturation
field H$_s$ is reached. For H $>$ H$_s$, the resistance rises in a
sublinear fashion. In the other bilayer, where CoPt is at the
bottom, the hysteresis in the resistivity is much more pronounced.

The M-H curves of the bilayers change dramatically on entering the
superconducting state as seen in Fig. 5. These data were taken at
5 K with in-plane field. For the CoPt-NbN/MgO bilayer, the M-H
curve is dominated by the diamagnetic response of the
superconducting NbN layer, showing a hysteresis loop typical of a
type-II superconductor. However, for the NbN-CoPt/MgO bilayer, the
ferromagnetic component of magnetization is distinctly seen. The
sudden jump in magnetization at $\pm$ 500 Oe coincides exactly
with the coercive field (H$_c$) seen in the M-H loop taken at 20 K
(Fig. 4(b)). Since the moment of CoPt is not expected to change
below 20 K because of its large Curie temperature ($\approx$ 710
K)\cite{Kootte}, the true diamagnetic response of the
superconductor can be extracted by subtracting the 20 K data from
the 5 K data. The continuous M-H curves in Fig. 5 (a $\&$ b) are
the true diamagnetic response. We can use the Bean critical state
model \cite{Classen} to extract the screening critical current
density J$_c$ form these M-H data. At field H $\approx$ 550 Oe
which is larger than the coercive field of both type of bilayers,
the J$_c$ of CoPt-NbN/MgO system is larger by a factor of $\simeq$
3.5 compared to the J$_c$ of the NbN-CoPt/MgO heterostructure. The
large suppression of J$_c$ in the latter structure suggests a
granular character of its NbN layer. It is worth pointing out that
the actual flux density in these films of thickness ($\approx$ 50
nm) smaller than the London penetration depth ($\approx$ 200 - 250
nm) will be lower due to the dipolar field of the CoPt layer which
induces reverse flux. However, the presence of the ferromagnetic
layer will not affect the relative magnitude of J$_c$ in the two
cases as long as the field is greater than the coercive field.

Now we discuss how the superconductivity in a granular and a
homogeneous film is affected when it is placed in proximity of a
ferromagnet with in-plane magnetization through measurement of
transport critical current density and its temperature and angular
dependence. Fig. 6(a) presents the J$_c$ measured with a voltage
criterion of 10 $\mu$V/cm as a function of reduced temperature
(T/T$_c$) for both the bilayers and a single layer NbN film. Here
T$_c$ denotes the temperature corresponding to the zero resistance
state of the samples. For the bilayer sample CoPt-NbN/MgO, the
J$_c$ at T/T$_c$ $\geq$ 0.9 is same ($\approx$ 6.0$\times$10$^5$
A/cm$^2$) as that of a single layer NbN film and below T/T$_c$ =
0.9, it exceeds the limit of our measurement which is set by the
maximum output of our current source ($\sim$ 100 mA). For the
other bilayer film (NbN-CoPt/MgO) however, the J$_c$ is highly
suppressed. For example, at T/T$_c$ = 0.8 it is only $\approx$
1.5$\times$10$^5$ A/cm$^2$. Fig. 6 also shows the J$_c$ vs.
T/T$_c$ plots of the two bilayers when a 300 Oe field is applied
perpendicular to the plane of the film. We note that the
field-induced suppression of J$_c$ is marginally higher in the
case of NbN-CoPt/MgO heterostructure. The temperature dependence
of J$_c$ is generally expressed by a phenomenal expression of the
type,
\begin{eqnarray}
J_c = J_0 (1 - T/T_c )^\beta, \label{eq1}
\end{eqnarray}
where J$_0$ and $\beta$ are used as fitting parameters. In the
Ginzburg-Landau (GL) mean-field description of J$_c$, the
prefactor J$_0$ is a measure of the depairing current and the
exponent $\beta$ is 3/2 \cite{Clem}. For a highly granular system
where superconducting grains are separated by insulating material,
the Ambegaokar-Baratoff model can be used to describe the J$_c$
(T) data \cite{Clem}. Here J$_0$ is related to superconducting gap
parameter and the exponent $\beta$ $\approx$ 1 at low
temperatures. The data shown in Fig. 6 have been fitted to Eq. 1.
The J$_0$ and $\beta$ for the NbN/MgO, CoPt-NbN/MgO and
NbN-CoPt/MgO in zero field are ($\approx$ 7.81$\times$10$^6$
A/cm$^2$, 0.93), ($\approx $1.65$\times$10$^7$ A/cm$^2$, 1.19) and
($\approx$ 1.64$\times$10$^6$ A/cm$^2$, 1.40) respectively. For
the first two samples, since the fitting has been done over a very
limited range of T/T$_c$, too much significance can not be given
to J$_0$ and $\beta$.  However, for NbN-CoPt/MgO excellent fitting
is seen for T/T$_c$ ranging from 0.5 to $\approx$ 1. It is
interesting to note that while the NbN in this sample has a
substantially reduced J$_c$, the $\beta$ remains close to the GL
value. The large value of $\beta$ here suggests a robust coupling
between NbN grains \cite{Clem}.

In order to examine how the state of magnetization of the CoPt
film affects superconductivity in NbN, we have measured the
temperature dependence of J$_c$ with in-plane field [Fig. 6(b)]
corresponding to points A and B of the hysteresis loop as shown in
the inset of Fig. 6(b). Since at point A of the loop the sample is
fully saturated, it should behave like a single domain magnetic
entity. Point B of the loop corresponds to a fully demagnetized
state where the sample consists of randomly oriented domains. The
direction of magnetization from one domain to the next can change
either by out-of-plane rotation of spins, which constitutes a
Bloch wall or by in-plane rotation as in the case of a
N$\acute{e}$el wall \cite{Robert}. Obviously, a Bloch wall will
lead to magnetic flux into the NbN layer where as for a
N$\acute{e}$el wall the flux remains confined in the ferromagnetic
film. A Bloch wall is therefore expected to reduce the J$_c$ of
the film. The J$_c$ of the NbN-CoPt/MgO and CoPt-NbN/MgO films
measured at positions A and B of the hysteresis loop is shown in
Fig. 6(b). No discernible difference in the critical current in
the two cases is seen suggesting that in these 50 nm CoPt films
the domain walls are of N$\acute{e}$el type.

In order to address further the likely perturbation of
superconductivity in NbN by magnetic domain structure of CoPt
layer, we show in Fig. 7 the angular dependence of flux flow
resistivity at three points of the hysteresis loop measured at a
temperature where resistance drops by $\approx$ 95 $\%$ of its
normal state value in zero-field. The orientation of current
$\vec{I}$, magnetization $\vec{M}$ and magnetic field $\vec{H}$
vectors are shown in the inset of the figure. The angle $\phi$ is
between $\vec{H}$ and film normal $\hat{n}$. At saturation points
A and D of the loop the $\vec{M}$ vector is always perpendicular
to $\vec{J}$, either along + $\hat{y}$ or - $\hat{y}$. The flux
flow resistivity of both bilayers at H = $\pm$ 1500 Oe (points A
and D of hysteresis) is characterized by two sharp cusps at $\phi$
= $\pm$ 90$^0$ and maximum dissipation at $\phi$ = 0$^0$, i.e.,
when the field is normal to the film surface. If we attribute the
flux flow resistivity only to the normal component of the field
(Hcos$\phi$), then $\rho$$_f$ must go linearly in Hcos$\phi$
following the relation for flux flow resistance
$\rho$$_f$/$\rho$$_N$ $\approx$ H/H$_{c2}$, where $\rho$$_N$ is
the normal state resistance and H$_{c2}$ the upper critical field
\cite{Tinkham}. In the right inset of Fig. 7 we show variation of
$\rho$$_f$ vs. Hcos$\phi$ for CoPt-NbN/MgO and NbN-CoPt/MgO
bilayers. In the former case, where superconductivity is robust in
NbN, the flux-flow resistance shows a linear dependence on
Hcos$\phi$. For the sample where NbN is on the top, however, the
scaling of $\rho$$_f$ with Hcos$\phi$ is poor, indicating that
this film can not be treated as an infinite superconducting plane.
In Fig. 7 we have also plotted the angular dependence of R$_f$ at
points B and C of the hysteresis loop. The large dissipation seen
at the coercive field in NbN-CoPt/MgO over a range of angles
around $\phi$ = 0 again points towards the granular nature of the
NbN.

\subsection{Bilayers with out-of-plane magnetic anisotropy}
Equiatomic CoPt films deposited at 700 $^0$C on MgO stabilize into
the L1$_0$ (tetragonal) ordered structure with a strong
out-of-plane anisotropy energy ($\approx$ 5$\times$10$^6$
J/m$^3$)\cite{McCurrie}. It is expected that the magnetic domain
structure of such films can lead to strong pinning of vortices in
a field range below the coercive field of the ferromagnet
\cite{Jan,Lange}. We have compared the relative influence of
L1$_0$ and A1 disordered CoPt layers on flux pinning in NbN in a
controlled experiment in which a 50 nm NbN film was first grown at
700 $^0$C on three MgO substrates mounted side-by-side in the
deposition chamber. On one of such films a 50 nm CoPt was
deposited at 700 $^0$C followed by annealing in high vacuum for 30
minutes while on the other the CoPt layer of the same thickness
was deposited at 600 $^0$C, but after annealing the NbN at 700
$^0$C for a time equal to the postannealing time used for L1$_0$
CoPt growth ($\approx$ 30 minutes). The third film was also
annealed for 40 minutes at 700 $^0$C under the identical
conditions to ensure the same metallurgical state of the NbN in
all three samples. In Fig. 8 we show superconducting transition of
these films measured resistively. While the T$_c$ onset of the
bare NbN is $\approx$ 15.5 K, a suppression of the critical
temperature by $\approx$ 0.2 K is seen on deposition of the
two-types of CoPt. This lowering of T$_c$ perhaps derives
contribution from exchange-field-induced pair-breaking in NbN.
From the X-ray diffraction pattern shown in inset `a' of Fig. 8,
it becomes clear that CoPt films grown at 600 and 700 $^0$C on NbN
are in the A1 (fcc) and L1$_0$ (tetragonal) polytypes with their
(001) direction normal to the plane of the film. Inset `b' of Fig.
8 shows magnetic hysteresis loop of these two bilayers measured at
20 K with the external field directed perpendicular to the film
plane. The square loop of the film deposited at 700 $^0$C clearly
indicates out-of-plane magnetic anisotropy axis with coercivity
$\approx$ 5 kOe. Having established the distinct magnetic and
crystallographic structure of the CoPt, we address the issue of
flux pinning in these bilayers. In Fig. 9 we compare the critical
current density of the NbN films with A1 and L1$_0$ CoPt capping
layer at three temperatures very close to T$_c$ (T/T$_c$ $\geq$
0.92) and in a field range H $\ll$ H$_c$ of the L1$_0$ film. The
magnetic field was directed perpendicular to the plane of the
film. It is very clear that the L1$_0$ CoPt-NbN couple has a small
but distinctly larger J$_c$ as compared to that of the A1 CoPt-NbN
in a field range H $<$ 1 kOe ($\ll$ H$_c$). To emphasize this
increase, in the inset of Fig. 9 we plot $\Delta$J$_c$ = [J$_c$
(L1$_0$) - J$_c$(A1)]. This enhancement of critical current can be
explained qualitatively in the framework of the model of
Bulaevskii, Chudnovsky and Maley (BCM) \cite{Bulkaevskii} which
shows that a flux line in superconducting film capped with a
magnetic layer of perpendicular anisotropy experiences a spatially
modulated pinning barrier of the type U$_{mp}$(x)$\sim$ $\Phi$$_0$
M(x) d$_s$, where $\Phi$$_0$ is the flux quantum, M(x), the
magnetization of the FM domain and d$_s$ the superconducting film
thickness. This behavior is relevant when the vortex motion is
perpendicular to domain wall, and the width of the domain $l$ is
large compared to d$_s$ and magnetic penetration depth
$\lambda$$_L$ $\ll$ $l$. Both these conditions are satisfied by
the CoPt film. The J$_c$ expected from such domain wall pinning is
$\sim$ c M$_0$/$l$; with M$_0$ $\approx$ 450 emu/cc for our L1$_0$
CoPt film, and $l$ $\approx$ 1000 nm, we get J$_c$ $\sim$
4.5$\times$10$^7$ A/cm$^2$ for vortex motion perpendicular to
domain walls. In real samples, however, the domains have arbitrary
geometry in which vortices gliding along domain walls will
experience no pinning force and thus a much smaller increase in
J$_c$ would result. For the A1 CoPt film, the perpendicular
component of magnetization at H = 1 kOe, where a detectable
enhancement in J$_c$ is seen, is very small and we do not expect
any magnetic pinning in this case.

In summary, we have studied the crystallographic structure,
magnetic ordering and superconducting properties of
ferromagnet-superconductor bilayers made of CoPt and NbN grown on
single crystal MgO in two different geometries, with either NbN or
CoPt in contact with the substrate. In the case of bilayers where
CoPt is deposited over NbN, a further distinction has been
realized by selecting ordered (L1$_0$) or disordered (A1)
polytypes of CoPt showing in-plane and perpendicular magnetic
anisotropy respectively. By comparing the SC response of bilayers
where the CoPt magnetization is in the plane of the structure with
that of a plane NbN film, we conclude that the ferromagnetism of
CoPt has no discernible effect on the bulk SC properties of the
NbN. This also leads us to believe that the magnetic domain walls
in A1 CoPt of both bilayers are of the N$\acute{e}$el type. While
the ferromagnetism of CoPt has no deleterious effects, the
polycrystalline nature of NbN deposited on CoPt imparts it a
granular character which reflects itself in a reduced T$_c$, in
the temperature dependence of J$_c$, and in the variation of
flux-flow resistance as a function of the angle between applied
field and film normal. A comparison of critical current density in
CoPt-NbN/MgO bilayers show a distinctly higher J$_c$ in samples
where the CoPt magnetization is out-of-plane. Since this gain is
seen only at fields smaller than the coercivity of the CoPt layer,
we attribute it to domain wall pinning of flux lines.

This research has been supported by grants from the Department of
Science $\&$ Technology under its Nanoscience $\&$ Nanotechnology
Initiative and by the Board for Research in Nuclear Science. S. K.
Bose acknowledges financial support from the Council for
Scientific and Industrial Research, Government of India. We also
acknowledge Mr. Pooran C. Joshi for his technical assistance.


\clearpage
\begin{figure}[h]
\vskip 0cm \abovecaptionskip 0cm
\includegraphics [width=12cm]{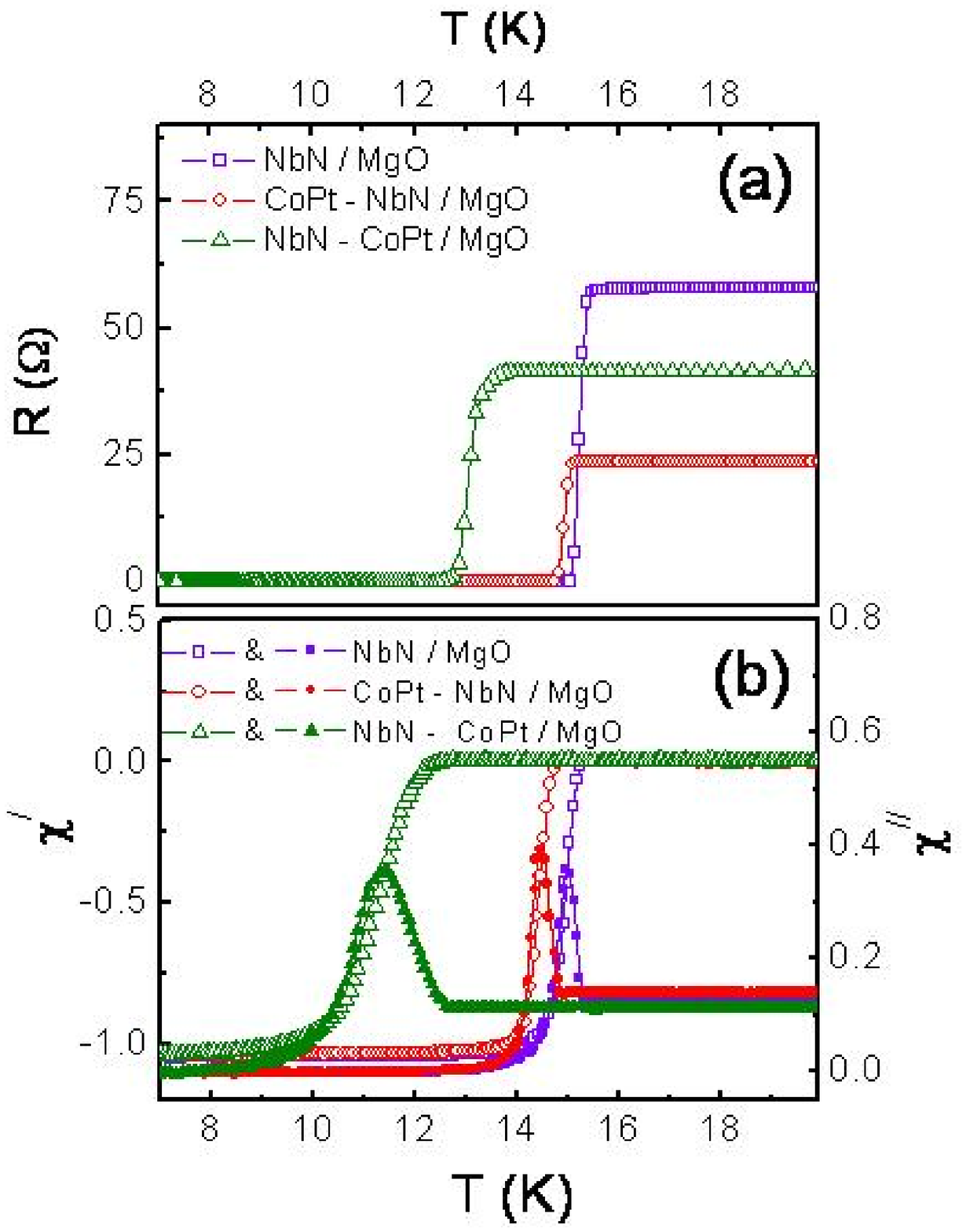}%
\caption{\label{fig1} Panel (a) shows the temperature dependence
of resistance for pure NbN, NbN-CoPt/MgO and CoPt-NbN/MgO bilayer
systems. $\chi^{'}$ and $\chi^{''}$  for all the three systems as
a function of temperature have been plotted in panel (b). }
\end{figure}

\clearpage
\begin{figure}[h]
\vskip 0cm \abovecaptionskip 0cm
\includegraphics [width=10cm]{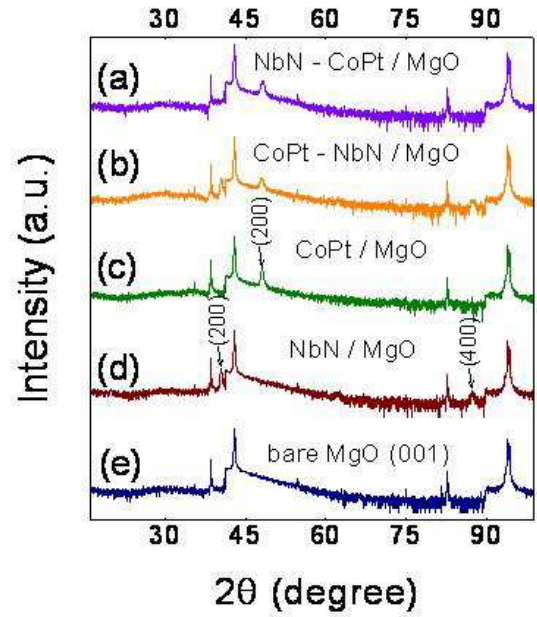}%
\caption{\label{fig2} (a) X-ray diffraction profiles collected
with a Seifert (model 3000P) diffractometer of (a) NbN-CoPt/MgO,
(b) CoPt-NbN/MgO, (c) CoPt, (d) NbN thin films deposited at 600
$^0$C on single crystal (001) cut MgO. The diffraction profile of
single crystal MgO (001) is also shown at the bottom panel. In
panel (c) and (d) peaks of A1 phase CoPt and rocksalt NbN
respectively are identified with miller indices.}
\end{figure}

\clearpage
\begin{figure}[h]
\vskip 0cm \abovecaptionskip 0cm
\includegraphics [width=12cm]{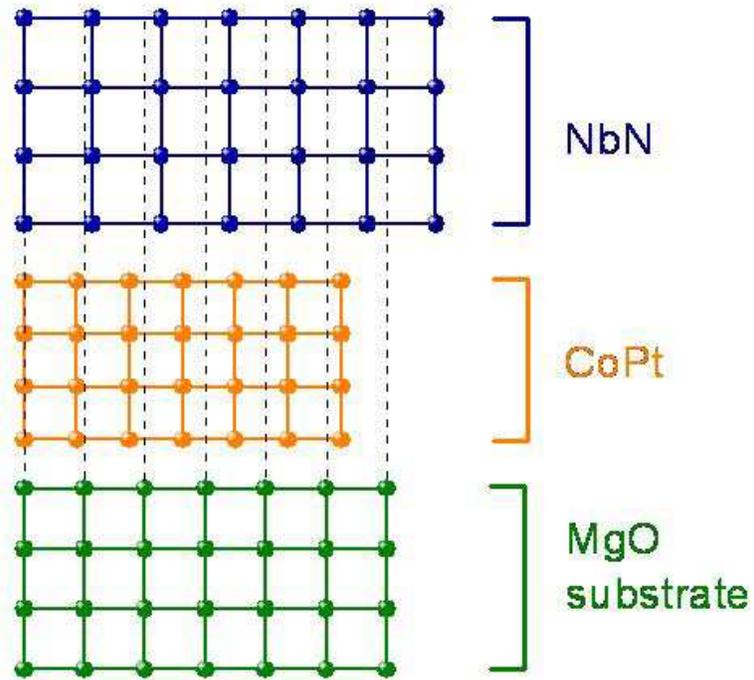}%
\caption{\label{fig3} Schematic illustration of strained
NbN-CoPt/MgO heterostructure. Dotted lines in the figure
represents ideal coherent epitaxy. Lattice mismatch is exaggerated
for clarity.}
\end{figure}

\clearpage
\begin{figure}[h]
\vskip 0cm \abovecaptionskip 0cm
\includegraphics [width=12cm]{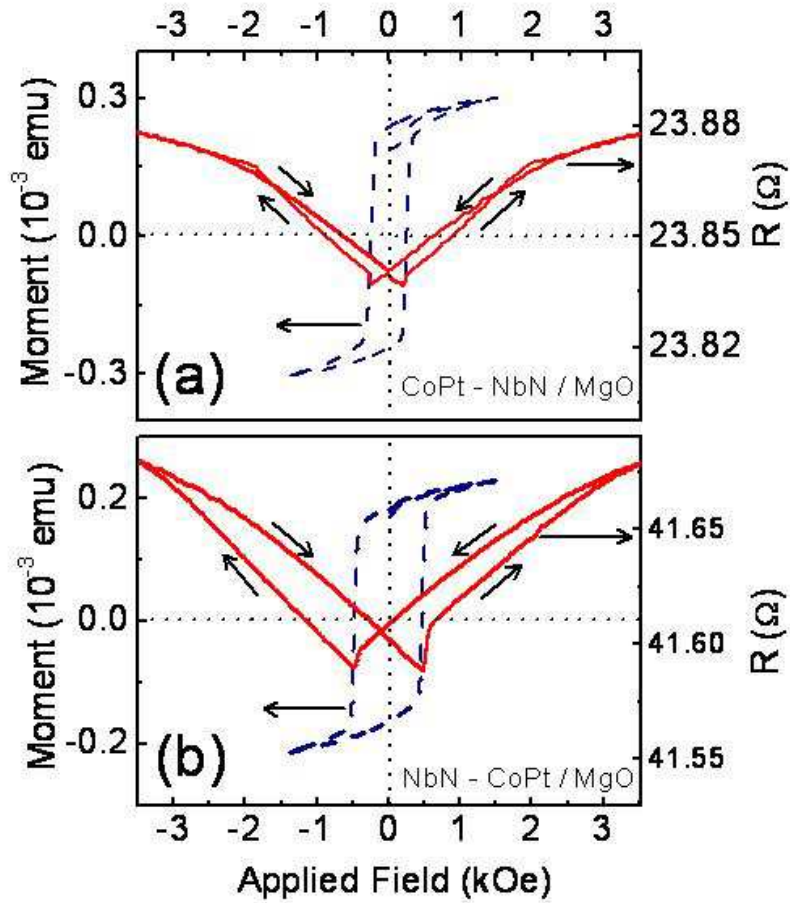}%
\caption{\label{fig4}Panel (a) shows isothermal magnetization and
resistance as a function of magnetic field applied in-plane for
CoPt-NbN/MgO system measured at 20 K. Panel (b) shows the
dependence of  magnetization and resistance on in-plane magnetic
field for NbN-CoPt/MgO system measured at 20 K. Arrows in the
figure mark the behavior of resistance on increasing and
decreasing field sweeps.}
\end{figure}

\clearpage
\begin{figure}[h]
\vskip 0cm \abovecaptionskip 0cm
\includegraphics [width=10cm]{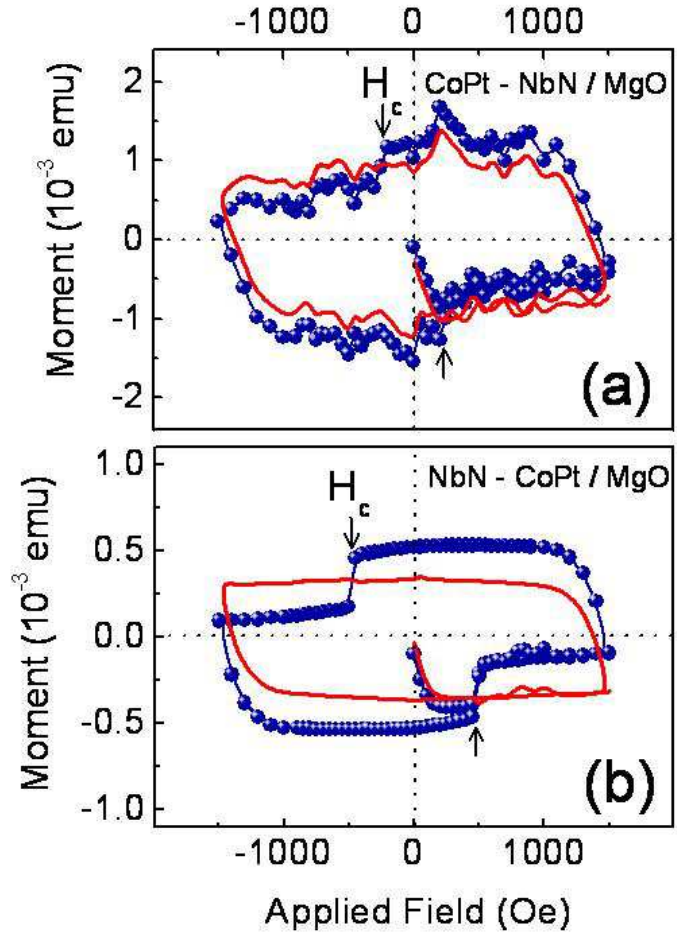}%
\caption{\label{fig5} Panels (a) and (b) show the in-plane
magnetization as a function of applied magnetic field measured at
5 K for CoPt-NbN/MgO and NbN-CoPt/MgO system respectively.
Contribution of superconducting NbN to total magnetization is
extracted by subtracting 5 K data from 20 K data and is shown in
the figures as continuous line.}
\end{figure}

\clearpage
\begin{figure}[h]
\vskip 0cm \abovecaptionskip 0cm
\includegraphics [width=10cm]{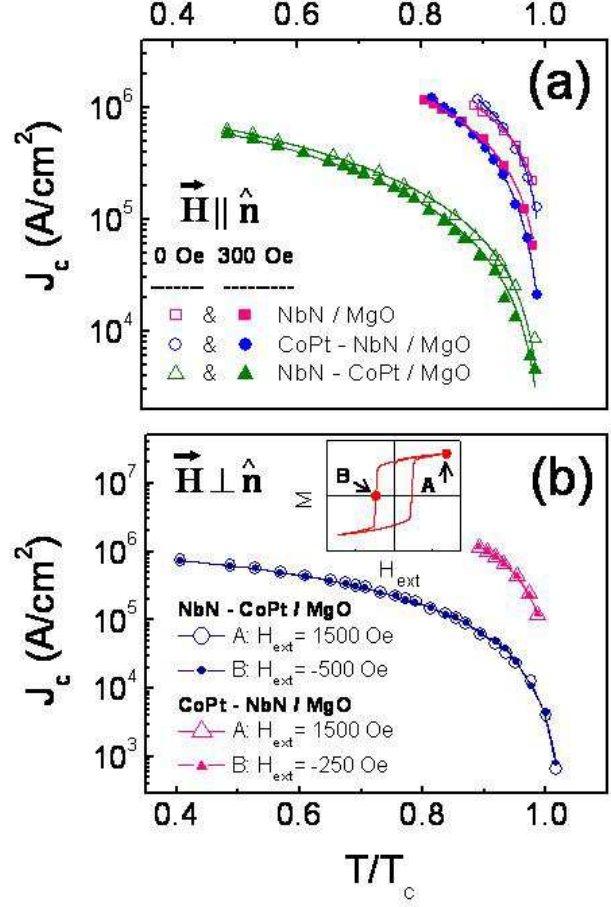}%
\caption{\label{fig6}(a) Plot of critical current density J$_c$
calculated from current-voltage characteristics using a field
criterion of 10 $\mu$V/cm of NbN, CoPt-NbN/MgO and NbN-CoPt/MgO as
a function of reduced temperature. Open symbols and solid symbols
represent the data for zero-field and 300 Oe perpendicular field
($\vec{H}$ $\parallel$ $\hat{n}$) measurements respectively, where
$\hat{n}$ is unit vector normal to the plane of the film. (b)
Temperature dependence of J$_c$ with in-plane magnetic field for
both the bilayers. In the inset letters A and B mark the points on
a typical M-H loop at which the measurements of J$_c$ were carried
out. While open symbols in panel (b) represent data taken at point
A, J$_c$ values for fully demagnetized state (point B) are plotted
as solid symbols.}
\end{figure}

\clearpage
\begin{figure}[h]
\vskip 0cm \abovecaptionskip 0cm
\includegraphics [width=16cm]{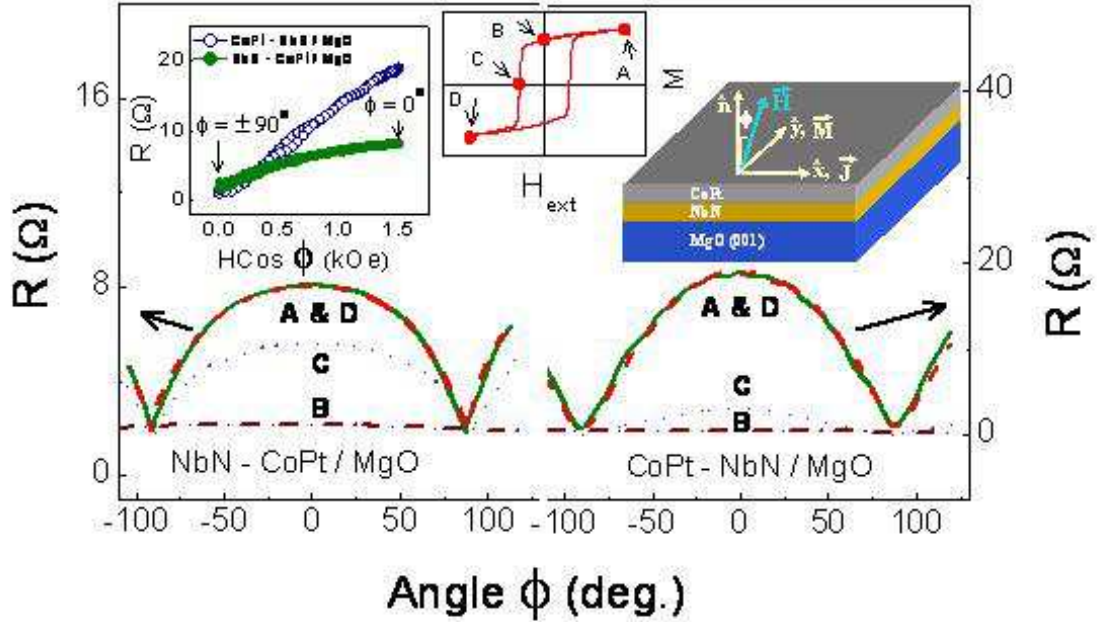}%
\caption{\label{fig7} The resistance of both CoPt-NbN/MgO (right
Y-axis) and NbN-CoPt/MgO (left Y-axis) systems in the flux flow
regime as a function of the angle $\phi$ between the applied
magnetic field and film normal ($\hat{n}$) at a temperature in the
transition region, where the sample resistance is 5 $\%$ of the
zero-field normal state resistance. Magnetic field direction was
changed in a step of 2 degree while it remained always orthogonal
to the current direction ($\vec{H}$ $\perp$ $\vec{J}$). Zeroes on
the X-axis correspond to field angle normal to the film surface. A
constant bias current of 1 $\mu$A was used for all the
measurements. Right inset shows schematically the direction of
current flow ($\vec{J}$), magnetization vector ($\vec{M}$) and the
direction of magnetic field ($\vec{H}$) which makes angle $\phi$
with $\hat{n}$. The letters A, B, C and D in the middle panel mark
the points on a typical M-H loop at which R was measured. Left
inset shows the dependence of resistance extracted from the data
presented in the main panel (curves A and D) as a function of
normal component of the applied field for both the samples.}
\end{figure}

\clearpage
\begin{figure}[h]
\vskip 0cm \abovecaptionskip 0cm
\includegraphics [width=12cm]{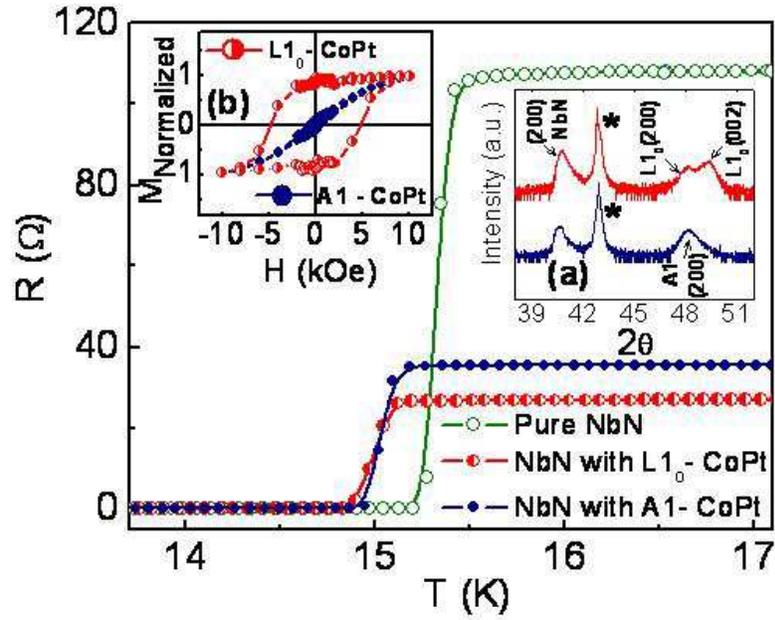}%
\caption{\label{fig8}Temperature dependence of resistance of
CoPt-NbN/MgO bilayers with both A1 and L1$_0$ phase CoPt is
compared with that of pure NbN film. Unlike the R(T) plot
presented in Fig. 1(a), here the NbN layer was deposited at 700
$^0$C for all the three systems. X-ray diffraction profiles
collected with X'Pert PRO MPD diffractometer of the same set of
bilayer samples are shown in the inset (a). Peaks of rocksalt NbN,
and A1 and L1$_0$ phase CoPt are identified with miller indices.
The diffraction peak arises from the substrate is marked with
asterisk. Inset (b) shows the isothermal magnetization with
perpendicular magnetic field measured at 20 K for both the
bilayers. }
\end{figure}

\clearpage
\begin{figure}[h]
\vskip 0cm \abovecaptionskip 0cm
\includegraphics [width=12cm]{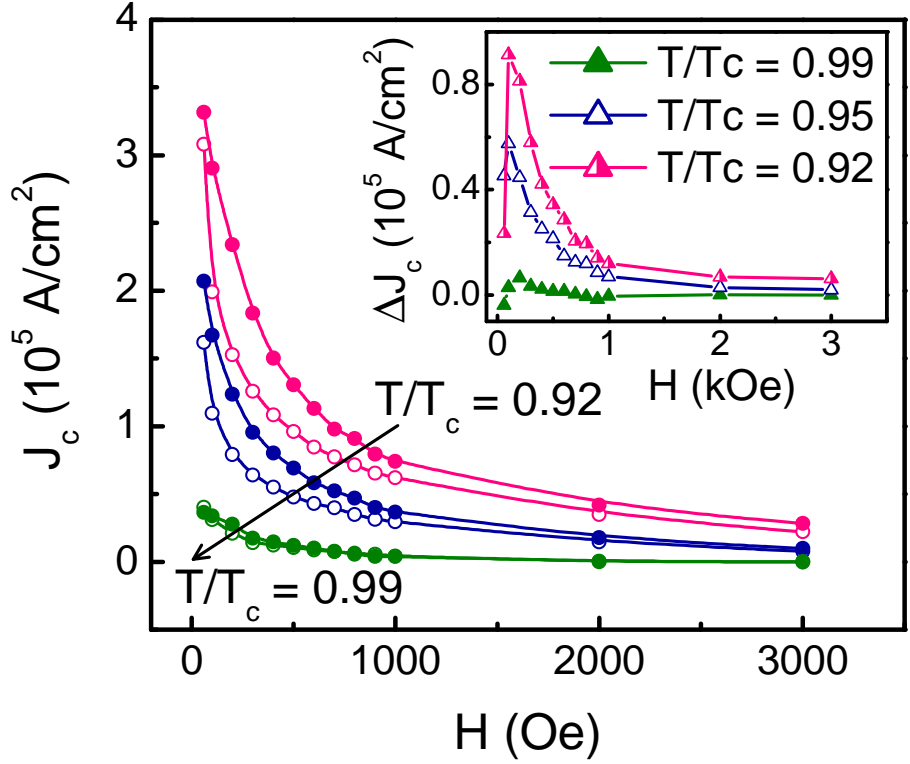}%
\caption{\label{fig9} Plot of critical current density J$_c$ of
(L1$_0$)-CoPt-NbN/MgO (solid symbols) and (A1)-CoPt-NbN/MgO (open
symbols) at three different temperatures (T/T$_c$ = 0.99, 0.95 and
0.92) as a function of magnetic field. The magnetic field was
directed perpendicular to the plane of the film. The first data
point corresponds to 60 Oe remanent field of the electromagnet.
Inset shows the plot of $\Delta$J$_c$ [= J$_c$ (L1$_0$) -
J$_c$(A1)] derived from the data shown in the main panel. }
\end{figure}

\end{document}